\documentclass[aps,amssymb,pra,twocolumn,superscriptaddress,showpacs,preprintnumbers]{revtex4}
\usepackage{epsfig}
\usepackage{bm}

\begin{document}


\title{Vortex signatures in annular Bose-Einstein condensates}
\author{M.~Cozzini}
\affiliation{Dipartimento di Fisica, Universit\`a di Trento and BEC-INFM, 
 I-38050 Povo, Italy}
\author{B.~Jackson}
\affiliation{Dipartimento di Fisica, Universit\`a di Trento and BEC-INFM, 
 I-38050 Povo, Italy}
\affiliation{School of Mathematics and Statistics, Merz Court, University of 
 Newcastle upon Tyne, NE1 7RU, United Kingdom}
\author{S.~Stringari}
\affiliation{Dipartimento di Fisica, Universit\`a di Trento and BEC-INFM, 
 I-38050 Povo, Italy}

\begin{abstract}
 We consider a Bose-Einstein condensate confined in a ``Mexican hat'' 
 potential, with a quartic minus quadratic radial dependence. We find
 conditions under which the ground state is annular in shape, with a hole in
 the center of the condensate. Rotation leads to the appearance of stable 
 multiply-quantized vortices, giving rise to a superfluid flow around the 
 ring. The collective modes of the system are explored both numerically and
 analytically using the Gross-Pitaevskii and hydrodynamic equations. Potential
 experimental schemes to detect vorticity are proposed and evaluated, which
 include measuring the splitting of collective mode frequencies, observing
 expansion following release from the trap, and probing the momentum
 distribution of the condensate.  
\end{abstract}
\date{\today}


\maketitle

\section{Introduction}

The 1995 discovery of Bose-Einstein condensation in dilute ultracold gases
\cite{anderson95,bradley95,davis95} prompted a flurry of activity investigating
these fascinating quantum systems. In the experiments, ultracold atoms are
confined by magnetic or optical forces, which typically can be represented
theoretically by a harmonic potential. Lately, however, an increasing amount of
interest has focused on Bose condensates in more exotic trapping potentials.
For example, Gupta {\it et al.} \cite{gupta05} recently produced a condensate
in a ring-shaped magnetic waveguide, while experiments at Ecole Normale
Sup\'{e}rieure \cite{bretin04,stock04} have been conducted in overlapping
magnetic and optical dipole traps, leading to a potential which can be
approximated by a function which is quadratic plus quartic in the radial
coordinate. This trap potential is especially of interest when the condensate 
is rotating, since, in contrast to the harmonic case, one can in principle
achieve arbitrarily high angular velocities. This leads to interesting new
configurations, for example a vortex lattice with a hole at its center, and for
even higher angular velocities, annular condensates containing a
multiply-quantized ``macrovortex''. These  equilibrium configurations have been
studied in a number of papers \cite{fetter01,lundh02,kasamatsu02}, where the
angular velocities for transitions between the different states, defining a
phase diagram, were calculated in
Refs.~\cite{kavoulakis03,jackson04,fetter05,kim05,fu05}. In addition to the
static properties, the dynamics have also been addressed, with the collective
mode excitations considered in Refs.~\cite{kim05,cozzini05}.

Alternatively, one could consider the case where, in addition to the quartic
term, there is a {\it negative} harmonic term. Then the radial trap potential
has the form of a ``Mexican hat'', and the condensate can be annular even when
there is no rotation. Similar trap configurations have been studied elsewhere
\cite{salasnich99,tempere01,nugent03,aftalion04}. Of particular interest,
however, is the behavior of the system under rotation, since vortices can enter
the condensate and form a stable, multiply-quantized vortex at the center. For
example, an important question raised in this configuration concerns the
stability of the vortex in the presence of a non-rotating thermal cloud at
finite temperatures--- does the vortex decay by spiraling towards the
edge as in a harmonic trap \cite{fedichev99,rosenbusch02}, or is it 
metastable?
This is related to the well-known problem of stability of persistent currents
in superfluids \cite{kagan00,bhattacherjee04}, and so is of interest 
from a fundamental perspective. 

In order to tackle these types of questions experimentally it is important to 
possess some way of detecting the presence of a vortex.
In a harmonically-trapped condensate this is relatively straightforward, since
the fluid circulation around the vortex core creates a
hole in the density, which can be detected by standard 
absorption imaging. Since the optical resolution is usually not sufficient 
to detect vortices {\it in situ}, a period of free expansion following the
release of the condensate from the trap is generally required.
However, in a trapped annular condensate vortices
reside in the center where the density is very low
irrespective of the circulation.
Hence, imaging of a vortex {\it in situ} is 
precluded, and it is not even obvious {\it a priori} whether it is possible to 
resolve the vortex after
expansion. A more detailed analysis is thus required in order to evaluate the
feasibility of this and other possible schemes for vortex detection.

In this paper we shall study two-dimensional condensates, where the trap
potential has a quartic minus quadratic radial dependence. Most
of our results for thin annular condensates, however, also apply to more
general Mexican hat potentials. Two dimensions in
this context corresponds to either a cylindrical condensate, where the axial
length scale greatly exceeds the radial size, or a thin disk-shaped condensate
where the system is tightly trapped in the axial direction such that dynamics
along this coordinate are ``frozen out'' and can be neglected.
Both geometries are of experimental interest. We also focus exclusively
on condensates in the mean field regime at zero temperature, where the 
properties can be accurately described by the Gross-Pitaevskii (GP) equation.
Nevertheless, we should mention that the possibility of obtaining a thin 
annulus, essentially a one-dimensional system, opens up other interesting 
regimes beyond mean field, such as a quasi-condensate or Tonks-Girardeau gas 
\cite{petrov00}.

We begin by studying the equilibrium states, where both  numerical solutions
and analytical results are  presented.
We then characterize the lowest energy excitations, or collective modes, of 
the condensate. In particular, we use a hydrodynamic model to derive 
analytical expressions in the limit of thin annuli, as well as to find the
mode frequencies numerically. We also compare these results to numerical
simulations of the GP equation.
This allows us to discuss the use of collective
modes to detect vorticity; namely the splitting between counter-propagating
surface and acoustic modes of the condensate.
We then consider the expansion of the condensate after switching off the
confining potential, finding that the hole due to the vortex 
is recovered in the expanded density, providing a simple means to detect the 
presence of the vortex in experiments.
Finally, we discuss the momentum distribution of an annular
condensate, finding that in the limit of thin annuli the function approaches a
simple analytical form.

\section{Equilibrium configurations}

For a dilute Bose-Einstein condensate in the 
limit of zero temperature, the time-dependent evolution of the 
condensate wavefunction $\Psi(\bm{r},t)$ is described by the Gross-Pitaevskii
(GP) equation
\begin{equation}
 i \hbar \frac{\partial \Psi}{\partial t} = \left( -\frac{\hbar^2}{2M}
 \nabla^2 + V_{\rm ext} + g_{\rm 3D} |\Psi|^2 \right) \Psi \, .
\label{eq:GP-3D}
\end{equation}
The interactions between atoms are represented by the parameter
$g_{\rm 3D}=4\pi N \hbar^2 a/M$, for $N$ atoms of mass $M$, with $s$-wave 
scattering length $a$.
Note that, in addition to the dynamics of the condensate, the 
time-independent stationary states can be found by substituting
$\Psi(\bm{r},t)=\Psi(\bm{r}){e}^{-i\mu t/\hbar}$
into Eq.~(\ref{eq:GP-3D}), whereupon the left-hand side becomes $\mu\psi$
with $\mu$ the chemical potential. 
The external potential in Eq.~(\ref{eq:GP-3D}) is of the form
\begin{equation}
 V_{\rm ext} (\bm{r})= \frac{\hbar \omega_{\perp}}{2} \left (\sigma \frac{r^2}
 {d_{\perp}^2} + \lambda \frac{r^4}{d_{\perp}^4} \right) + 
 \frac{\hbar \omega_z}{2} \frac{z^2}{d_z^2} \, ,
\label{eq:pot}  
\end{equation}
where $d_i=\sqrt{\hbar/(M\omega_i)}$, with $\omega_{\perp}$ and $\omega_z$
the radial and axial trap frequencies respectively. In the following we 
shall mostly use harmonic oscillator units where distance,
time and energy are expressed in units of $d_{\perp}$, $\omega_{\perp}^{-1}$
and $\hbar \omega_{\perp}$ respectively. We also take $\sigma=-1$, where the 
case of $\sigma=1$ has already been extensively studied in a number of papers 
\cite{fetter01,lundh02,kasamatsu02,kavoulakis03,jackson04,fetter05,kim05,fu05,cozzini05}. 

We explicitly consider the GP equation in 2D
\begin{equation}
 i \frac{\partial}{\partial t}  \Psi = \left[ -\frac{1}{2}\nabla^2 +
   \frac{1}{2} \left( -r^2+\lambda r^4 \right) + g |\Psi|^2 \right] \Psi \, ,
\label{eq:GP}
\end{equation}
where the evolution in the axial coordinate is neglected.
Physically this can correspond to the case of a cylindrical condensate, 
where $\omega_z \ll \omega_{\perp}$. The interaction parameter then becomes,
$g=4\pi N' a$, in harmonic oscillator units, where $N'=N/Z$ represents the 
number of atoms per unit length along the axial direction. An alternative 
two-dimensional configuration involves tight axial confinement, such that
$\hbar\omega_z \gg g_{3D}|\Psi(0)|^2$ in physical units.
In this geometry, which has been realized experimentally \cite{gorlitz01}, the
axial dynamics are ``frozen out''. The axial dependence of the  wavefunction is
then gaussian, and the 2D GP equation~(\ref{eq:GP}) can be used with
$N'=N/(\sqrt{2\pi} d_z)$ \cite{pitaevskii}. 

We first analyze a condensate without a vortex using the Thomas-Fermi (TF)
approximation, where the spatial derivative of $\Psi$ in
Eq.~(\ref{eq:GP}) is neglected, so that the density is given by 
$n(\bm{r})\equiv|\Psi(\bm{r})|^2=(\mu-V_{\rm ext})/g$ for 
$n(\bm{r})>0$, and $n(\bm{r})=0$ otherwise. The radii of the condensate are
evaluated by finding where the density goes to
zero. If $\mu>0$ then there is a single solution of the resulting quadratic
equation, $R^2=(1+\sqrt{1+8\lambda\mu})/(2\lambda)$, which corresponds to a 
condensate without a hole. On the other hand, if $\mu<0$ there are two 
solutions $R_{1,2}^2=(1\mp\sqrt{1+8\lambda\mu})/(2\lambda)$, and the 
condensate is annular with a hole in the center. In this case one can write the
TF condensate density as
\begin{equation}
 n(\bm{r}) = \frac{\lambda}{2g} (R_2^2-r^2)(r^2-R_1^2) \, ,
\label{eq:density}
\end{equation}
for $n(\bm{r})>0$.
Calculating the radii and chemical potential more 
explicitly requires the
condition that the wavefunction is normalized to unity $\int {\rm d}\bm{r} \,
n(\bm{r}) = 1$. The resulting radii can be conveniently expressed as the sum
and difference of the squares $R_{\pm}^2=R_2^2 \pm R_1^2$, yielding
\begin{equation}
 \lambda R_+^2 = 1 \, ,
\label{eq:Rp}
\end{equation}
\begin{equation}
 \lambda R_-^2 = \eta \equiv \left( \frac{12 g \lambda^2}{\pi} 
 \right)^{1/3} \, ,
\label{eq:Rm}
\end{equation}              
where we have defined a new parameter $\eta$ in Eq.~(\ref{eq:Rm}). The 
chemical potential is then
\begin{equation}
 \mu = \frac{\eta^2-1}{8\lambda} \, . 
\label{eq:mu}
\end{equation}
This yields the condition $\eta<1$ for an annular condensate,
which we will always assume in the following.
The limit $\eta\to0$ corresponds to the case where the width of the annulus
becomes much smaller than the radius.

The range of validity of the TF approximation for an annular condensate is
defined by the condition $\xi \ll d$ \cite{fetter05}, where 
$\xi=(2gn_{\rm max})^{-1/2}$ is the 
healing length at the density maximum, while $d=R_2-R_1$ is the width of
the annulus. Using the above expressions yields the explicit relations
$\xi=2\sqrt{\lambda}/\eta$ and
$d=(\sqrt{1+\eta}-\sqrt{1-\eta})/\sqrt{2\lambda}$.

The condensate can also contain a single centered vortex 
with circulation $\Gamma=2\pi\nu\hbar/M$, where the well-known requirement of
quantized circulation in superfluids \cite{donnelly} constrains the
vorticity to integer values $\nu=0,1,2,...\, $.  
The equilibrium condensate density is represented within the TF
approximation as
\begin{equation}
 g n(\bm{r}) = \mu - \frac{1}{2} \left( \frac{\nu^2}{r^2} 
 - r^2 + \lambda r^4 \right ) \, ,
\label{eq:density-nu}
\end{equation}
where $\mu$ is the chemical potential in the laboratory (non-rotating) frame
and the $\nu^2/r^2$ term corresponds to the centrifugal potential barrier
resulting from the fluid flow around the vortex. Performing an analysis 
similar to that of the non-rotating condensate gives the 
following relations
\begin{equation}
 (\lambda \nu)^2 = \frac{(\lambda R_+^2 -1)[(\lambda R_+^2)^2 - 
 (\lambda R_-^2)^2]}{4} \, , 
\label{eq:lambnu-rot}
\end{equation}
\begin{equation}
 \lambda \mu = \frac{1}{2} \left[ \frac{3(\lambda R_+^2)^2 + 
 (\lambda R_-^2)^2}{4} - \lambda R_+^2 \right] \, ,
\label{eq:lambmu-rot}
\end{equation}
\begin{equation}
 \frac{\eta^3}{12} = \frac{2}{3} \lambda^2 \mu R_-^2 -
 \frac{(\lambda \nu)^2}{2} \ln \left(\frac{R_+^2+R_-^2}
 {R_+^2-R_-^2} 
 \right) + \frac{\lambda^2 R_+^2 R_-^2}{12} \, .
\label{eq:eta-rot}
\end{equation}

Approximate analytical solutions of
Eqs.~(\ref{eq:lambnu-rot}-\ref{eq:eta-rot}) can be obtained using the
expansion $\lambda R_+^2 = 1 + \sum_n c_n  (\lambda \nu)^{2n}$, which is most
appropriate for $\lambda\nu \ll 1$. As we shall see this regime is the most
interesting for our purposes, since it corresponds to the region where a
multiply-quantized vortex is stable. Substituting into
Eqs.~(\ref{eq:lambnu-rot}-\ref{eq:eta-rot}) yields the following results
\begin{equation}
 \lambda R_+^2 = 1 + \frac{4}{1-\eta^2} (\lambda \nu)^2 + \mathcal{O} 
 [(\lambda \nu)^4 ] \, ,
\label{eq:exp-rp}
\end{equation}
\begin{eqnarray}
 \lambda R_-^2 = \eta+\frac{1}{\eta} \left[ \frac{4}
 {\eta^2-1}+\frac{2}
 {\eta} \ln \left( \frac{1+\eta}{1-\eta} \right) \right] (\lambda
 \nu)^2 \\ \nonumber
 + \mathcal{O}[(\lambda \nu)^4 ] \, ,
\label{eq:exp-rn}
\end{eqnarray}
\begin{equation}
 \lambda \mu = \frac{\eta^2-1}{8} +  \frac{1}{2\eta} \ln \left(
 \frac{1+\eta}{1-\eta} \right)
 (\lambda \nu)^2 + \mathcal{O} [(\lambda \nu)^4] \, .
\label{eq:exp-mu}
\end{equation}

Another important quantity is the energy of the condensate in the presence of
a vortex. This is represented in terms of an energy per 
particle as an integral
\begin{equation}
 E = \frac{1}{2}\int {\rm d} \bm{r} \ \Psi^* \left( - \nabla^2 -r^2 +\lambda 
 r^4 + gn \right) \Psi \, .
\label{eq:energy-GP}
\end{equation}
Within the TF approximation the energy becomes
\begin{equation}
 E = \frac{1}{2}\int {\rm d} \bm{r} \, \left( \frac{\nu^2}{r^2} -r^2 
 +\lambda r^4 + gn \right) n \, .
\label{eq:energy-TF}
\end{equation}
Eq.~(\ref{eq:energy-TF}) can be approximated analytically using the TF 
density~(\ref{eq:density-nu})
and the expansions~(\ref{eq:exp-rp}-\ref{eq:exp-mu}). This eventually yields
\begin{equation}
 \lambda E = -\frac{1}{8} + \frac{3}{40} \eta^2 + E_1 (\lambda \nu)^2  + 
 \mathcal{O} [(\lambda \nu)^4] \, ,
\label{eq:energy-ana1}
\end{equation}
where we have defined the second-order term separately
\begin{equation}
 E_1 = \frac{3}{4\eta^3}
 \left[ 2\eta - (1-\eta^2) \ln \left( \frac{1+\eta}{1-\eta} \right) 
 \right] \, .
\label{eq:e_1}
\end{equation}
Interestingly, this term can also be derived by calculating the kinetic energy 
from the circulation of the vortex $\int {\rm d}\bm{r}\, n \nu^2/(2r^2)$, where
$n$ is approximated by the density without a vortex (\ref{eq:density}).

The energy in a frame rotating with angular velocity $\Omega$ is related to 
that in the laboratory frame by $E'(\nu)=E(\nu)-\Omega\nu$. For a particular 
$\Omega$ there exists a vorticity $\nu$ that minimizes this energy, identifying
the ground state for each rotation rate. Treating $\nu$ as a
continuous
quantity (valid for $\nu \gg 1$) and using Eq.~(\ref{eq:energy-ana1}) to
second order gives the following condition for the ground state vorticity
\begin{equation}
 \lambda \nu = \frac{\Omega}{2 E_1} \, .
\label{eq:nu}
\end{equation}
In the limit $\eta \to 0$ this expression becomes particularly simple, 
since $E_1 \simeq 1$, so that $\lambda \nu \simeq \Omega/2$. This can
also be easily derived by equating the velocity due to the rotation at the 
radius of the annulus, $v=\Omega R$, to the fluid velocity due to the 
vortex, $v=\nu/R$, where $R^2=R_+^2/2=1/(2\lambda)$.
  
These analytical results can be compared to solutions of the GP equation in 
the rotating frame
\begin{equation}
 i \frac{\partial}{\partial t}  \Psi = \left[ -\frac{1}{2}\nabla^2 +
   \frac{1}{2} \left( -r^2+\lambda r^4 \right) + g |\Psi|^2  -
   \Omega \hat{L}_z \right] \Psi \, ,
\label{eq:GP-rot}
\end{equation}
where $\hat{L}_z = i (y\partial/\partial x-x \partial/\partial y)$ is the axial
component of the angular momentum operator. To find the  equilibrium
configuration for each $\Omega$ we solve Eq.~(\ref{eq:GP-rot}) numerically
by propagating in imaginary time (i.e.\ making the replacement $t \to -it$),
starting from the $\Omega=0$ ground state.

\begin{figure}[here]
\centering \scalebox{1.0}
 {\includegraphics{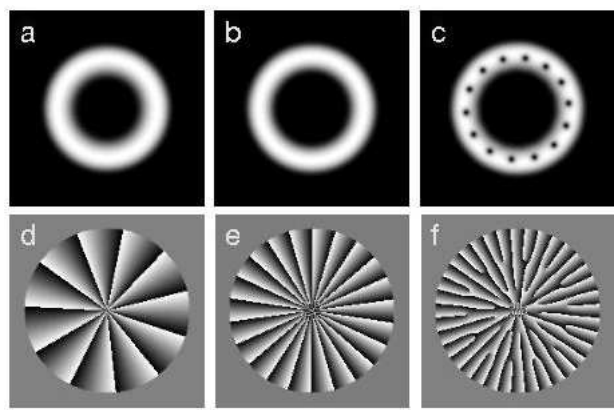}}
 \caption{Density profiles of a condensate, found by numerical solution of the
 GP equation (\ref{eq:GP-rot}) with $g=1000$ and $\lambda=0.01$
 ($\eta=0.726$) for (a) $\Omega=0.2$ (in units of $\omega_{\perp}$), 
(b) $\Omega=0.4$, and (c)
 $\Omega=0.5$; (d)-(f) show the corresponding phase profiles for each
 $\Omega$. The length scale of each box is $28 \times 28$ (in units of
 $d_{\perp}$).}
\label{fig:deph}
\end{figure}

This technique yields the configurations shown in Fig.~\ref{fig:deph}. One can
write the condensate wavefunction as $\Psi(\bm{r}) = \sqrt{n(\bm{r})} \exp
[iS(\bm{r})]$, where  $n(\bm{r})=|\Psi(\bm{r})|^2$ is the density and
$S(\bm{r})$ is the phase. Fig.~\ref{fig:deph} shows the density and phase at
three different rotation rates for a particular set of parameters ($g=1000$,
$\lambda=0.01$, $\eta=0.726$). For $\Omega=0.2$ (Fig.~\ref{fig:deph}(a)) the
equilibrium state consists of a stable multiply-quantized vortex with $\nu=11$,
as can be seen from plotting the phase (Fig.~\ref{fig:deph}(d)) where, by
traversing the center in a closed loop, one finds 11 cycles between $S=0$
(black) and $S=2\pi$ (white). At $\Omega=0.4$ (Figs.~\ref{fig:deph}(b) and (e))
the multiply-quantized vortex is again stable, but with $\nu=22$. However, at
$\Omega=0.5$ (Figs.~\ref{fig:deph}(c) and (f)) a multiply-quantized vortex at
the center of the condensate is surrounded by two rings of singly-quantized
vortices. The outer ring is visible 
in the density plot since the vortices are embedded 
in the condensate. This transition between a 
``macrovortex'' and a
``lattice with hole'' state was previously found in Ref.~\cite{aftalion04} for
3D configurations. This is similar 
to the transition in quartic plus quadratic traps 
discussed in Refs.\ \cite{kavoulakis03,jackson04,fetter05,kim05,fu05},
except that in the latter case the transition takes place at a much higher
frequency and in the opposite direction. 
To find the critical rotation rate for the transition in the present 
configuration would require a detailed analysis, possibly using 
techniques similar to Refs.\ \cite{kim05,fu05}.
   
\begin{figure}[here]
\centering \scalebox{0.45}
 {\includegraphics{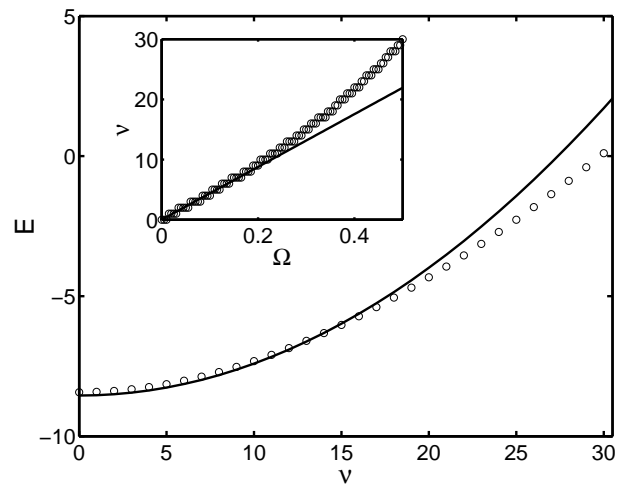}}
 \caption{Energy (in units of $\hbar\omega_{\perp}$) of the condensate with a
 central vortex as a function of vorticity $\nu$ for the same parameters as
 Fig.~\ref{fig:deph}. The circles plot numerical solutions of the GP equation,
 while the solid line is the analytical estimate (\ref{eq:energy-ana1}). {\it
 Inset:} vorticity of the ground state in a frame rotating with angular
 velocity $\Omega$ (in units of $\omega_{\perp}$). Circles are numerical GP
 solutions, while the solid line plots the analytical result (\ref{eq:nu}).}
\label{fig:energyvort}
\end{figure}

In the remainder of the paper we shall focus on the macrovortex state,
(i.e.\ one where there is a single, multiply-quantized vortex in the 
center of the hole) where in all cases we have checked that
the macrovortex is stable in the rotating frame by numerically solving 
the GP equation. The analytical results derived above can then be used, and
compared to numerical solutions of the GP equation. In order to facilitate such
a comparison, it is simpler to solve the GP equation in the non-rotating frame
with an effective potential 
$V_{\rm ext}(r,z=0)+\nu^2/(2r^2)$. The energy can then be
calculated from the density using Eq.~(\ref{eq:energy-GP}) and is
compared to the analytical expression (\ref{eq:energy-ana1}) in
Fig.~\ref{fig:energyvort}. One sees good agreement for $\nu<20$
(corresponding to $\lambda \nu < 0.2$),
which is the region where Eq.~(\ref{eq:energy-ana1}) is expected to be valid.
To find the ground state for a particular rotation rate $\Omega$ one can,
as before, simply
identify the $\nu$ that minimizes the rotating frame energy
$E'(\nu)=E(\nu)-\nu\Omega$. The resulting $\nu$ {\it vs.\ }$\Omega$ plot is
compared to the analytical result (\ref{eq:nu}) in the inset of
Fig.~\ref{fig:energyvort}, again showing good agreement for small 
$\lambda\nu$. 

Before continuing, it is interesting to relate our work to more
general ``Mexican hat'' type potentials--- for example, a harmonic plus
gaussian potential \cite{nugent03} which experimentally would correspond
to magnetic trap overlaid with a tightly-focussed blue-detuned laser beam. In
particular, for a thin annulus, when the condensate is confined close to 
the potential minimum at radius $R$, the radial trap potential can be 
expanded to second order in $r-R$ as 
\begin{equation}
 V(r)=\frac{M}{2}\omega_r^2 (r-R)^2 + \mathcal{O} [(r-R)^3] \, \,
\label{eq:potexp}
\end{equation}
where we have neglected a constant term which does not affect the 
dynamics. For our potential (\ref{eq:pot}) we find 
$\omega_r=\sqrt{2}\,\omega_\perp$ and 
$R=d_{\perp}/\sqrt{2\lambda}$ (in physical units). In the following, 
the results presented for the thin annulus limit $\eta \ll 1$ are 
therefore applicable to more general potentials, where finding 
$\omega_r$ and $R$ from the corresponding expansion 
(\ref{eq:potexp}) allows a precise mapping onto our system.

\section{Collective modes}

In the previous section it was found that, within the TF approximation, a hole 
is present within a non-rotating condensate for $\eta<1$. Considering a
condensate rotating with sufficiently small angular velocities $\Omega$, we 
found equilibrium configurations where multiply-quantized vortices are 
stable. An interesting
question concerns how these vortices could be detected experimentally, since
the density profiles of a condensate with no vortex or small values of $\nu$
are very similar, making it very difficult to distinguish between the different
states on this basis alone. One possibility is to study the collective modes of
the system, since differences in the vorticity may be manifested in the
frequencies, in a similar way to those in a harmonic trap
\cite{zambelli98,chevy00,haljan01}. The problem of collective modes in annular 
condensates is also interesting in its own right, as similar investigations in
a rapidly rotating quadratic plus quartic trap have shown \cite{cozzini05}.
This section will therefore detail similar analyses for a quartic minus
quadratic confinement, with the specific application to vortex detection
discussed at the end.

One approach to finding collective mode frequencies is to rewrite the GP 
equation
in the rotating frame (\ref{eq:GP-rot}) in terms of the density and phase, and
use the TF approximation to obtain  hydrodynamic equations for these
quantities. Then linearizing for small fluctuations about equilibrium,
$\delta{S}$, $\delta{n}$, yields
\begin{eqnarray}
  \frac{\partial}{\partial t}\,\delta{n}+
  \left(\frac{\nu}{r^2}-\Omega\right)\frac{\partial\delta{n}}
  {\partial\phi}+\bm\nabla\cdot(n_0\bm\nabla\delta{S}) = 0 \, ,
  \label{eq:HD lin irn} \\
  \frac{\partial}{\partial t}\,\delta{S}+
  \left(\frac{\nu}{r^2}-\Omega\right)\frac{\partial\delta{S}}
  {\partial\phi}+g\delta{n} = 0 \, , \label{eq:HD lin irs}
\end{eqnarray}
where $n_0$ is the TF equilibrium density from  Eq.~(\ref{eq:density-nu}). For
density and phase modulations of the form  $\delta{n},\,\delta{S}\propto
{e}^{i(m\phi-\omega t)}$ this gives
$\delta{n}=i(\omega+m\Omega-m\nu/r^2)\delta{S}$ and
\begin{eqnarray}
 \left[\left(\omega+m\Omega-\frac{m\nu}{r^2}\right)^2-\frac{m^2}{r^2}
 gn_0\right]\delta{S} \hspace{2cm} \nonumber \\
 +\frac{1}{r}\frac{\partial}{\partial r}\left(rgn_0 
 \frac{\partial}{\partial r}\delta{S}\right) = 0 \, ,
\label{eq:hydro}
\end{eqnarray}
where $\omega$ is the frequency of the mode in the rotating frame.

It is instructive to first consider the case of a non-rotating condensate 
where $\Omega=0$ and $\nu=0$. Similarly to Ref.~\cite{cozzini05}, one can
introduce the new variable $\zeta=(r^2-R_+^2/2)/(R_-^2/2)$, so that the
equilibrium density (\ref{eq:density}) becomes $gn_0 = \lambda R_-^4
(1-\zeta^2)/8$ and $-1\leq\zeta\leq1$. Substituting into Eq.~(\ref{eq:hydro})
then gives
\begin{eqnarray} \label{eq:HD lin no rot}
 \left(\omega^2-\frac{m^2\eta^2}{4}\frac{1-\zeta^2}{1+\eta\zeta}
 \right)\delta{S} \hspace{3.5cm} \nonumber \\
 +\frac{\partial}{\partial\zeta}\left[(1+\eta\zeta)
 (1-\zeta^2)\frac{\partial}{\partial\zeta}\delta{S}\right] = 0 \, .
\label{eq:hydro-zeta}
\end{eqnarray}
Note that this equation, and therefore the mode frequency, depends only upon
the parameter $\eta$ and the square of the azimuthal quantum number $m$, the
positive and negative $m$ modes being degenerate for symmetry reasons. In the
limit of $\eta \to 0$ (i.e.\ when the width of the annulus is small compared to
its radius, as evident from the relation $\eta=R_-^2/R_+^2$) the equation
becomes
\begin{equation}
 \omega^2\delta{S}+\frac{\partial}{\partial\zeta}\left[(1-\zeta^2)
 \frac{\partial}{\partial\zeta}\delta{S}\right] = 0 \, ,
\end{equation}
which is Legendre's equation with eigenfunctions given by Legendre polynomials
$P_j(\zeta)$ and eigenvalues $\omega^2=j(j+1)$, where $j=0,1,2,\dots$ is the
number of radial nodes \cite{spurious}. Hence, different $m$ states are
degenerate in this limit \cite{corrections}. The same
result for the frequencies can
also be derived starting from the expanded potential (\ref{eq:potexp}) with 
the corresponding one-dimensional hydrodynamic equation for the radial
coordinate. This implies that
the modes can be pictured as consisting of different parts of the 
annular condensate undergoing radial oscillations
in a one-dimensional harmonic well, providing a simple explanation for
the mode frequencies and the degeneracy of different $m$ states.

\begin{figure}[here]
\centering \scalebox{0.57}
 {\includegraphics{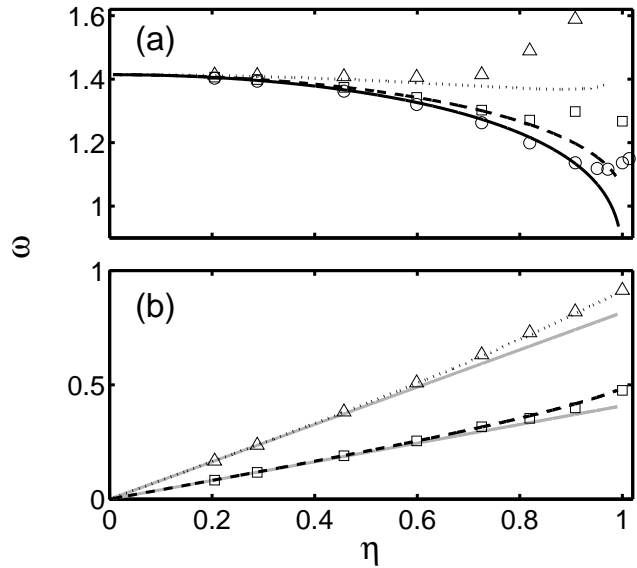}}
 \caption{(a) Frequency (in units of $\omega_{\perp}$) of the
 high-lying ($j=1$) $m=0$  (solid line) $m=1$ (dashed) and
 $m=2$ (dotted) modes as a function of $\eta$ for a non-rotating condensate
 ($\Omega=0$), from numerical solution of the  hydrodynamic equation
 (\ref{eq:hydro-zeta}). Also plotted are 
 numerical solutions of the Gross-Pitaevskii
 equation (\ref{eq:GP}) at $g=1000$ for $m=0$ ($\circ$), $m=1$ ($\square$) and
 $m=2$ ($\triangle$). (b) Frequencies of the low-lying
 ($j=0$) $m=1$ and $m=2$ modes, where the grey lines show the analytical
 result (\ref{eq:lowfreq}) for each $m$, while the other lines and points are
 labelled similarly to (a).} 
\label{fig:freq1}
\end{figure}

Apart from these analytical considerations, one can also solve
Eq.~(\ref{eq:hydro-zeta}) numerically using a 
shooting method \cite{nr} to find the
mode frequencies $\omega$ and functions $\delta S(\zeta)$.
Fig.~\ref{fig:freq1}(a) shows the high-lying ($j=1$)
$m=0,1,2$ mode frequencies as a function of $\eta$ for the non-rotating
($\Omega=0$, $\nu=0$) condensate in the presence of the hole ($\eta<1$). One
sees that in the limit of $\eta \to 0$ all of the mode frequencies tend to
$\omega = \sqrt{2}$, consistent with the above analysis. With
increasing $\eta$, however, the frequencies deviate from this value, with a
particularly marked decrease in the $m=0$  mode as one approaches $\eta=1$. In
addition, it becomes increasingly difficult to find numerical solutions near to
$\eta=1$. Numerical results from  simulating the time-dependent GP equation
(\ref{eq:GP}) for each mode are also presented in Fig.~\ref{fig:freq1}(a). The
GP results closely follow the hydrodynamic equations up to $\eta=0.7$, above
which the GP results start to deviate significantly. 
This discrepancy is related to the fact that when the inner radius approaches 
zero at $\eta \to 1$ the contrast between the density profiles for the TF
approximation (where the density goes to zero suddenly) and from solving the GP
equation (where the density tails off more gradually) becomes increasingly
significant, leading to differences in the nature of the modes near to the
center. In addition, the transition between hole and non-hole states in the GP
solution is more gradual, avoiding the singular behavior in the hydrodynamic
equation at $\eta=1$. 
 
The low-lying ($j=0$) $m>0$ modes can also be investigated,
and the $m=1$ and $m=2$ modes frequencies are plotted as a function of 
$\eta$ in 
Fig.~\ref{fig:freq1}(b). Comparison between the hydrodynamic and GP results
again show very good agreement. Similar modes were found in a rapidly rotating
quadratic plus quartic traps in Ref.~\cite{cozzini05}, and in the thin  annulus
limit $\eta \ll 1$ correspond to compressional modes directed  azimuthally
around the annulus, in contrast to the high-lying modes which correspond to
shape oscillations of the annulus. This physical explanation is the basis of a
simple analysis, starting from the relation $\omega=cq$, where for a narrow
annulus $q=m/R$, with the mean radius $R=R_+/\sqrt{2}$. The speed of sound, 
$c$, is given by the relation $c^2=n_1 \partial \mu / \partial n_1$,  where
$n_1=N'/(2\pi)$ is the density of the condensate integrated over the radial
direction. Rewriting in terms of $\eta$ yields the equation
\begin{equation}
 \omega^2=\frac{2m^2}{3\lambda R_+^2} \eta \frac{\partial}{\partial \eta}
 (\lambda \mu) \, .
\label{eq:lowfreq-gen}
\end{equation}
Substituting Eqs.~(\ref{eq:Rp}) and~(\ref{eq:mu}) for a non-rotating 
condensate, $\lambda \nu=0$, gives $c=\eta/\sqrt{12\lambda}$ and
\begin{equation}
 \omega=|m| \frac{\eta}{\sqrt{6}} \, .
\label{eq:lowfreq}
\end{equation}
The mode frequencies for $m=1$ and $m=2$ found from
Eq.~(\ref{eq:lowfreq}) are plotted as a function of $\eta$ in
Fig.~\ref{fig:freq1}(b).
This analytical estimate is exact for thin annuli \cite{corrections},
and indeed the comparison to the hydrodynamical results shows excellent
agreement for $\eta<0.4$.

So far we have only studied the non-rotating case where $\Omega=0$. In a frame
rotating with a non-zero angular frequency, $\Omega \neq 0$, we demonstrated
earlier that for sufficiently small $\Omega$ the ground state consists of a
multiply quantized vortex $\nu>1$. These stationary states can be described
analytically within the TF approximation using Eq.~(\ref{eq:density-nu})
for the density. From the normalization condition for the density one can find
the chemical potential and energy for each $\nu$, followed by the $\nu$ state
which minimizes the energy in a frame rotating with frequency $\Omega$.
Substituting the resulting $\nu$ and $n_0$ into Eq.~(\ref{eq:hydro}) and
solving numerically then yields mode frequencies for this particular $\Omega$.

We find that the rotating-frame frequencies for the high-lying ($j=1$)
modes change significantly with increasing $\Omega$, with a rather large
splitting between $m=+|m|$ and $m=-|m|$ modes for a given $|m|>0$. In contrast,
the $\Omega$ dependence of the low-lying ($j=0$) mode frequencies is
less pronounced, and is almost independent of the sign of $m$. 
In the laboratory frame, however, one has a frequency splitting 
between low-lying modes of opposite $m$. Using a sum rule analysis similar
to that of Ref.~\cite{zambelli98} gives for this splitting
\begin{equation}
 \Delta \omega = 4|m|\lambda \nu [1+\mathcal{O}(\eta^2)]\, ,
\label{eq:splitting}
\end{equation}
where a similar expression was derived using 
perturbation theory in Ref.~\cite{nugent03} for a thin toroid in a quadratic 
plus gaussian potential. For the thin annulus ($\eta \ll 1$) one can 
use Eqs.~(\ref{eq:lowfreq}) and (\ref{eq:splitting}) to derive a simple 
equation for the frequencies of the low-lying modes in the laboratory frame
\begin{equation}
 \omega_{\pm} = |m|\left( \frac{\eta}{\sqrt{6}} \pm 2\lambda \nu \right) \, .
\label{eq:lowfreq-rot}
\end{equation}
Note that these expressions can also be derived by using the 
quasi-degeneracy between modes in the rotating frame, together with the 
transformation $\omega_{\rm lab}=\omega_{\rm rot}+m\Omega$ and 
Eq.~(\ref{eq:nu}) with $E_1\simeq 1$.

\begin{figure}[here]
\centering \scalebox{0.53}
 {\includegraphics{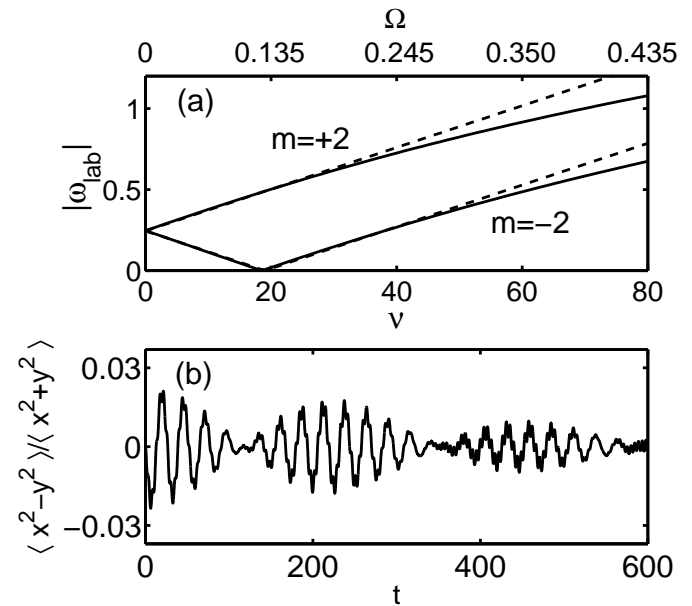}}
 \caption{ 
 (a) Frequencies in the laboratory frame of the low-lying $m=\pm2$ 
 modes as a function of $\nu$ and $\Omega$, for $g=680$ and
 $\lambda=0.00322$. The solid lines corresponds to numerical solutions
 of Eq.~(\ref{eq:hydro}), while the dashed lines show the analytical 
 expression
 (\ref{eq:lowfreq-rot}) for both modes. (b) The quadrupole moment
 $\langle{x^2-y^2}\rangle$ (normalized to the mean squared radius 
 $\langle{x^2+y^2}\rangle$) as a function of time 
 (in units of $\omega_{\perp}^{-1}$) given by a numerical 
 simulation of the GP equation in the laboratory
 frame. The original condensate contains a single $\nu=1$ vortex, and the
 applied perturbation mainly excites the low-lying $m=\pm 2$ modes.} 
\label{fig:freq5}
\end{figure}

In Fig.~\ref{fig:freq5}(a) we plot the laboratory-frame frequencies,
$|\omega_{\rm lab}|$, of the $m=\pm2$ low-lying modes as a function of $\nu$
and $\Omega$ for a thin annulus, where $\lambda=0.00322$, $g=680$, and 
$\eta=0.3$. To place
these values into physical context, note that they would correspond to e.g.\
$10^6$ $^{87} {\rm Rb}$ atoms in a cylindrical condensate $100\, \mu {\rm
m}$ in length, with trap frequency $\omega_{\perp}/(2\pi) \simeq 20 {\rm Hz}$
giving a length unit of $d_{\perp}\simeq 2.4\, \mu{\rm m}$. This would give an
annulus with inner and outer radii of $25\, \mu{\rm m}$ and $34\, \mu{\rm m}$
respectively, and is of interest as it corresponds to feasible parameters in 
future experiments. Eq.~(\ref{eq:lowfreq-rot}) is also plotted in
Fig.~\ref{fig:freq5}(a), where we see very good agreement with the
hydrodynamic results for $\nu<40$.

Above a particular value of $\nu$ we find that $\omega$ of the $m=-2$ mode
becomes negative. From Eq.~(\ref{eq:lowfreq-rot}) this  critical value is
found to be independent of $|m|$ and given by $\nu_{\rm crit} =
\eta/(\lambda\sqrt{24})$, in good agreement with the numerical hydrodynamic
results. This corresponds to the point at which the $m=-|m|$ modes 
become energetically
unstable, whereupon a source of dissipation (such as the presence of a thermal
cloud at non-zero temperatures) results in the mode amplitudes increasing with
time. 
This can also be seen from the famous Landau criterion \cite{pitaevskii},
which states that excitations in a superfluid become energetically unstable at
flow velocities, $v$, exceeding the speed of sound, $c$. For a thin annulus
surrounding a vortex the fluid velocity is $v=\nu/R$, while
$c=\eta/\sqrt{12\lambda}$, immediately leading to the equation for $\nu_{\rm
crit}$ derived above.

Another interesting consequence of the fact that the mode frequencies 
pass through zero is related to a possible experiment where one applies a 
small, static perturbation of multipolar form to a condensate containing a 
$\nu \simeq \nu_{\rm crit}$ vortex. This is 
similar to the experiment performed for a vortex lattice in a 
harmonically-trapped condensate \cite{engels02}, whereupon the negative 
$m$ modes are resonantly excited and the resulting non-linear behavior 
studied.

An important question relates to the feasibility of exploiting the splitting
between the positive and negative mode frequencies as a diagnostic for the
presence of a vortex, in analogy with the similar use of the surface mode
splitting in harmonically-trapped condensates
\cite{zambelli98,chevy00,haljan01}. We test this idea in more detail by
performing numerical simulations based on the time-dependent GP equation.
We start with a condensate with the above parameters, and with a $\nu=1$ vortex
in the center. A perturbation is added to the lab.\ frame trap potential of
the form $\delta{V}=\alpha(x^2-y^2)$ up to $t=1$, after which the potential is
returned to its original form. The subsequent oscillations are tracked by
calculating various moments of the system $\langle \chi \rangle (t)= \int {\rm
d}\bm{r}\, |\Psi(\bm{r},t)|^2 \chi(\bm{r})$ as a function of time. 

The $\langle{x^2-y^2}\rangle$ quadrupole moment (normalized to the mean
squared radius $\langle x^2+y^2 \rangle$) is shown in
Fig.~\ref{fig:freq5}(b) for $\alpha=0.0025$. The perturbation predominately
excites the low-lying $m=+2$ and $m=-2$ modes \cite{perturbation}, and analysis
of the time series shows that modes with (lab.\ frame) frequencies
$\omega=0.230$ and $\omega=0.257$ are present, which agree well with the
predictions from the analytical estimate (\ref{eq:lowfreq-rot}) of
$\omega_-=0.232$ and $\omega_+=0.258$. Importantly the splittings also
closely agree, within 5\,\%, allowing Eq.~(\ref{eq:lowfreq-rot}) to be used to
deduce $\nu$ from measurements of the splitting. Physically the frequency
splitting is responsible for the ``beating'' between the two modes evident in
the plot in Fig.~\ref{fig:freq5}(b).
In experiments these modes would be apparent as a density pattern around the 
annulus, consisting of four antinodes with alternate maxima 
and minima. The presence of
the splitting would then lead to rotation of the position of the antinodes,
similar to the ``precession'' of the quadrupole surface modes seen
experimentally in harmonic traps 
\cite{chevy00,haljan01}. The 
precession frequency $\omega_{\rm pr}$ is equal to 
$\Delta \omega/(2|m|)$ \cite{pitaevskii}, corresponding to the angular 
velocity of the reference frame where the modes are degenerate. Note that
in the annular condensate the precession frequency is simply 
\begin{equation}
 \omega_{\rm pr} = 2\lambda\nu \simeq \Omega\, ,
\label{eq:prec-freq}
\end{equation}
independent of $|m|$, in contrast to 
the harmonically trapped case where the precession frequency depends upon
the multipolarity. The second equality is valid for large $\nu$ 
and small $\eta$, where one can use Eq.~(\ref{eq:nu}) with 
$E_1 = 1$. The simulations were repeated for the $|m|=1$ modes, as well
as the $|m|=2$ modes in the presence of a $\nu=2$ vortex, yielding 
frequencies that agree well with the predictions of 
Eq.~(\ref{eq:lowfreq-rot}).

Thus, the splitting of the modes may be observable and should match closely 
with that found from simple analytical estimates. However, a limiting factor 
could be
the ability of experiments to resolve the density fluctuations associated with
the mode, since this would probably be more difficult than observing shape
changes as in harmonic traps. One could in principle increase the amplitude of
the mode by using a large perturbation (large $\alpha$), but as seen in
Fig.~\ref{fig:freq5}(b) this can lead to nonlinear coupling of the mode to 
higher excitations, leading to a rapid damping (note that this damping is
absent for smaller perturbations, e.g.\ $\alpha=0.0005$). So there is a trade
off involved in choosing the size of the perturbation, and this may in
practice limit the utility of collective modes for vortex detection. 

\section{Expansion}

One alternative to using mode frequencies to detect vorticity in annular 
condensates was proposed by Tempere {\it et al.}~\cite{tempere01}, and 
involves allowing an expanding condensate to overlap with an another expanding
condensate without a vortex. The resulting interference pattern reveals the
presence of a vortex in the first condensate. However, this method may be
difficult to realize experimentally, and a much simpler procedure would be to
use only one condensate and allow it to expand freely after switching off the
trapping potential. The central question to be addressed in this section is
whether imaging the resulting density profile would provide a means to
distinguish between states containing a vortex and those without.

\begin{figure}[top]
\centering \scalebox{0.48}
 {\includegraphics{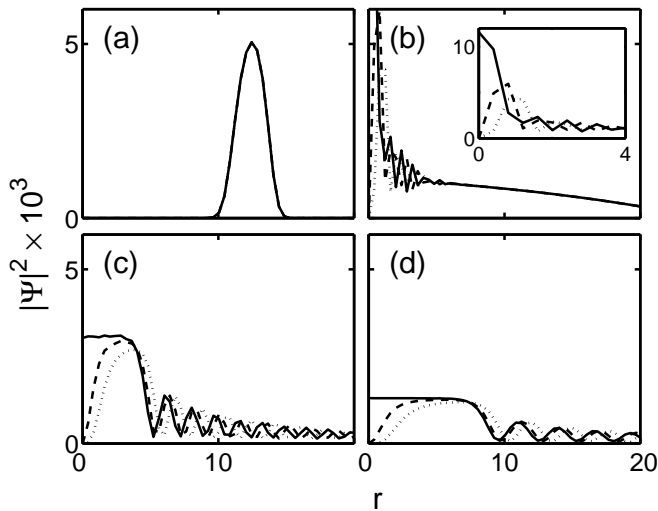}}
 \caption{Density as a function of radial distance through an expanding
 condensate ($\lambda=0.00322$, $\eta=0.3$) with no vortex (solid line), a
 $\nu=1$ singly-quantized vortex (dashed) and a $\nu=2$ doubly-quantized
 vortex (dotted) at (a) $t=0$, (b) $t=4$, (c) $t=8$, (d) $t=12$. The inset in
 (b) shows the density profiles near to the center, illustrating the
 differences between the states in this region.} 
\label{fig:exp-xsec}
\end{figure}

We study this question by starting with the GP equilibrium state either with or
without a vortex, then propagating the 2D time-dependent GP equation 
with $V_{\rm ext}$ set to zero. We study the narrow annulus case mentioned 
at the end of the
previous section, where $\lambda=0.00322$ and $\eta=0.3$.
Fig.~\ref{fig:exp-xsec} shows radial cross-sections through the density for
$\nu=0$, $\nu=1$ and $\nu=2$ for various times during the expansion.
Initially at $t=0$ (Fig.~\ref{fig:exp-xsec}(a)) there is very little difference
between the three configurations, illustrating the
difficulty in distinguishing between
the three states {\it in-situ}. After release from the trap ($t>0$) we see that
the condensate expands radially both outwards and inwards, thereby filling the
annulus. So, even at $t=4$ (Fig.~\ref{fig:exp-xsec}(b)) the different states
are, in principle, distinguishable due to the density hole produced by the
centrifugal barrier near to the vortex core. One can also see the
development of interference fringes away from the center. 

Figs.~\ref{fig:exp-xsec}(c) and (d) show the condensate after further 
expansion.
The size of the density dip at the vortex core is proportional to the healing
length, $\xi=1/\sqrt{2gn}$, so that the vortex expands along with the
condensate due to the corresponding decrease in density. The vortex size can be
quantified in the simulations by measuring the width at half the maximum 
density of the condensate, which for $\nu=1$ gives radii of 0.75 at $t=8$ and
$1.2$ at $t=12$. Using the example parameters introduced in the last section,
this would correspond to vortex core diameters of $3.6\, \mu{\rm m}$ and $5.6\,
\mu{\rm m}$ at $64\, {\rm ms}$ and $95\, {\rm ms}$ respectively, approaching
vortex sizes resolvable in experiments \cite{madison00}.

\begin{figure}[top]
\centering \scalebox{0.65}
 {\includegraphics{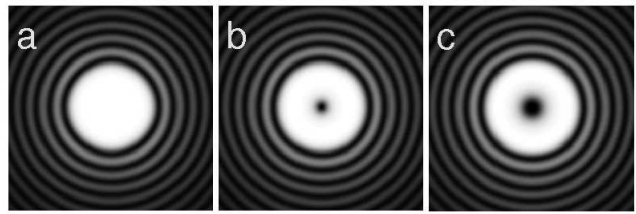}}
 \caption{Density of the expanding condensate ($\lambda=0.00322$, $\eta=0.3$)
 at $t=14$ with (a) no vortex  ($\nu=0$), (b) a singly-quantized vortex
 ($\nu=1$), and (c) a doubly-quantized vortex ($\nu=2$). The length scale of
 each box is $28 \times 28 \, d_{\perp}$.} 
\label{fig:threepan}
\end{figure}

Note that at later times the $\nu=0$ condensate density near to the center is
relatively uniform, which could provide a means to differentiate between
different $\nu$ states experimentally. For a uniform condensate the radius of
the vortex core at half-maximum density can be found numerically 
\cite{pitaevskii} yielding values of
$1.5\xi$ and $3.0\xi$
for $\nu=1$ and $\nu=2$. This factor of two difference is
reproduced in our simulations, which at $t=14$ yield vortex radii at
half-maximum of $1.4$ and $2.7$ for $\nu=1$ and $\nu=2$ respectively. Using the
predicted value $\xi=0.9$, from the measured peak density, yields good
estimates for the vortex radii. So in experiments, by measuring the vortex core
sizes and using simple theoretical relations, one could determine whether
either a singly or a doubly quantized vortex is present \cite{3D expansion}.
In general, the larger the vorticity, the wider will be 
the central hole at a given instant after release from the trap.
However, the quantitative measurement of large vorticity values could in
practice be limited by the longer expansion times needed to yield a wide enough
central plateau, the corresponding reduction in density 
making the experimental imaging difficult.

Fig.~\ref{fig:threepan} shows the 2D density profile at $t=14$ for $\nu=0$,
$\nu=1$ and $\nu=2$, which would be similar to the absorption images typically
found in experiments.
As well as showing the density profiles are circularly-symmetric, it also
illustrates the large differences between the images, which adds qualitative
support to the idea that it should be relatively simple to distinguish between
different vortex states after expansion.     

It is important to note that the central plateau is a consequence of
repulsive interatomic forces, and for free expansion with no interactions a
more sharply peaked distribution is found. In this latter case the condensate
wavefunction is given by the analytical expression \cite{pitaevskii}
\begin{equation}
 \Psi(\bm{r},t) = {e}^{-iD\pi/4}\!\left(\frac{M}{2\pi\hbar{t}}\right)^{\!D/2}
 \!\!\!\int\!\psi(\bm{r}',0){e}^{iM(\bm{r}-\bm{r}')^2/2\hbar{t}}\text{d}\bm{r}' \, ,
\label{eq:free-expand}
\end{equation}
where $D$ is the dimensionality, so in this case $D=2$. Note also that we have 
returned to using physical units. At large times,
$t \gg MR_+^2/\hbar$,
the density of the expanded condensate is circularly symmetric and tends
towards
\begin{equation}
 |\Psi (\bm{r}, t \to \infty)|^2 = \left( \frac{M}{\hbar t} \right)^2
 \left| \Phi \left( \frac{Mr}{\hbar t} \right) \right| ^2 \, ,
\label{eq:free-inft}
\end{equation}
where $|\Phi(k)|^2$ is the momentum distribution of the condensate. This will 
be discussed in more detail in the next section.

\section{Momentum distribution}

A final method for vortex detection involves considering the momentum 
distribution of the condensate, which could be probed experimentally by
switching off interactions during expansion (e.g.\ by exploiting a Feshbach
resonance) as discussed in the previous section, or by Bragg spectroscopy
{\it via} the dynamic structure factor \cite{stenger99,zambelli00}.
Mathematically, the momentum distribution $|\Phi(\bm{k})|^2$ is obtained
by Fourier transforming the equilibrium
wavefunction $\Psi(\bm{r})=|\Psi(r)| \,{e}^{i\nu\phi}$.  After integrating
over the angular coordinate (i.e.\ performing a Hankel transform) one obtains
\begin{equation}
 |\Phi(k)| = \int^{\infty}_0 {\rm d} r \, \, r |\Psi(r)| J_{\nu} (kr) \, ,
\label{eq:fourier-gen}
\end{equation}
where we have taken the modulus to remove constant phase factors, and
$J_{\nu}(kr)$ is the Bessel function of order $\nu$. As we have already seen,
for small $\nu$ one can approximate the equilibrium density in position space
with the TF profile (\ref{eq:density}). In this case the integral can be
rewritten in terms of the variable $\zeta=(r^2-R_+^2/2)/(R_-^2/2)$ as
\begin{equation}
 |\Phi(k)| = \sqrt{\frac{3\eta}{32\pi\lambda}} \int^1_{-1} {\rm d} \zeta
 \, \, \sqrt{1-\zeta^2} J_{\nu} \left(\frac{k \sqrt{1+\eta \zeta}}
 {\sqrt{2\lambda}} \right) \, . 
\label{eq:fourier-TF}
\end{equation}
The limit of this integral for a narrow annulus ($\eta \ll 1$) is of particular
interest. A series expansion of the Bessel function in powers of $\eta$ then
gives the simple result
\begin{equation}
 |\Phi(k)| = \sqrt{\frac{3\pi\eta}{128\lambda}} \left[J_{\nu} (kR) +
 \mathcal{O} (\eta^2) \right] \, ,
\label{eq:fourier-bess}
\end{equation}
where $R=1/\sqrt{2\lambda}$ is the radius of the annulus. Note that odd powers
of $\eta$ disappear due to parity and that the accuracy of this
expansion decreases for large $k$.
  
\begin{figure}[here]
\centering \scalebox{0.55}
 {\includegraphics{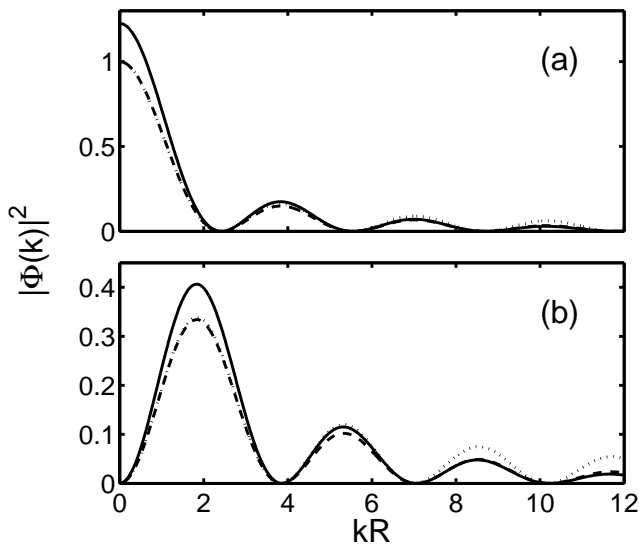}}
 \caption{Scaled momentum distributions $128\lambda|\Phi(k)|^2/(3\pi \eta)$ as
 a function of $kR$ (where $R=1/\sqrt{2\lambda}$) for (a) $\nu=0$, and (b)
 $\nu=1$ ($\lambda=0.00322$, $\eta=0.3$). Solid lines plot the (squared)
 integral (\ref{eq:fourier-gen}) calculated with solutions of the
 Gross-Pitaevskii equation, while the dashed lines show
 Eq.~(\ref{eq:fourier-TF}) for the Thomas-Fermi approximation. For comparison,
 the squared Bessel functions $[J_\nu (kR)]^2$ are plotted with dotted lines.} 
\label{fig:momdist}
\end{figure}

Fig.~\ref{fig:momdist} shows the cylindrically-symmetric momentum 
distributions $|\Phi(k)|^2$ for $\nu=0$ and $\nu=1$ and previously used
parameters $\lambda=0.00322$, $\eta=0.3$. The solid line corresponds to a 
numerical evaluation of Eq.~(\ref{eq:fourier-gen}), where the wavefunction is
found by solving the time-independent Gross-Pitaevskii equation, while
the dashed line plots the TF result (\ref{eq:fourier-TF}). These are scaled for
comparison to the Bessel function $J_{\nu}(kR)$. One sees that the Bessel
function closely approximates the TF result at small momenta, with differences
appearing at larger $kR$. Further calculations show that the agreement degrades
for larger $\eta$, as would be expected, although it
is still reasonably good for $kR<5$.

There is also a significant difference between the TF and full GP solutions,
mostly in the amplitudes of the peaks. This is despite the fact that 
$\xi \simeq 0.1\, d$
so the condition for the validity of the TF approximation is
well satisfied. Crucially, however, the positions of the minima
and maxima of the momentum distribution are relatively insensitive to the
details of the wavefunction. In particular, for both $\nu=0$ and $\nu=1$ the
first few zeros and maxima of $|\Phi(k)|^2$ from the GP solution
agree well with the roots of the corresponding Bessel functions and their
derivatives. By virtue of this close correspondence, the positions of these
peaks and troughs could provide a means to measure vorticity in experiments.
For example, the first maximum could be particularly useful, since its position
shifts to higher $k$ as $\nu$ increases. 

\section{Summary}

We have studied a dilute Bose-Einstein condensate in a ``Mexican hat'' trapping
potential of quartic minus quadratic form. We use the Thomas-Fermi (TF)
approximation to analytically find the equilibrium properties of both
non-rotating and rotating condensates, and compare to numerical solutions of
the Gross-Pitaevskii (GP) equation. In particular, we find that a non-rotating
condensate has an annular structure, with a hole in the center, for $\eta<1$,
where $\eta=(12g\lambda^2/\pi)^{1/3}$. We also find that for condensates in a 
frame rotating at sufficiently low angular velocities a stable 
multiply-quantized vortex with circulation $\nu>1$ is present at the center. 
At higher angular velocities our Gross-Pitaevskii simulations show that
the ground state can instead consist of a centered
multiply-quantized vortex surrounded by rings of singly-quantized vortices,
some of which are visible within the annulus.

We have also investigated the dynamics of a condensate with and without a
multiply-quantized vortex, particularly the collective modes. A hydrodynamic
equation was derived within the
TF approximation, with solutions that agree well with the results of
evolving the GP equation for large interaction strengths and
numbers of atoms where the TF approximation is expected to be valid. We find
that for a non-rotating condensate in the limit of a thin
annulus ($\eta\to0$) the mode frequencies are
$\omega=\sqrt{j(j+1)}$,
where the number of radial nodes $j=0,1,2,...$ defines families of modes 
containing different angular quantum numbers $m$. In this limit modes with the
same $j$ but different $m$ are
degenerate, but diverge as one increases $\eta$. The $j=0$ family for $m>0$ are
low-lying modes corresponding to sound waves directed around the annulus.
Thus a simple analytical expression for their frequencies can be derived from
$\omega=cq$, where $c=\eta/\sqrt{12\lambda}$ is the speed of sound, and
$q=|m|/R$ with $R=1/\sqrt{2\lambda}$ the radius of the annulus, giving 
$\omega=|m|\eta/\sqrt{6}$.

The mode frequencies in the presence of a multiply-quantized vortex were also
analyzed, where it was found that the frequency difference between opposite $m$ modes in the rotating frame is much smaller than the
splitting introduced when transforming to the laboratory frame, especially for
the low-lying modes. The possibility of using this splitting for experimental
detection of vorticity is explored, in particular the excitation of low-lying
$m=\pm 2$ modes in a thin annulus. 
Although in theory the splitting could be observable as a ``precession'' of the
mode around the annulus, the practicality of this technique may be limited due
to the difficulty of observing the associated small density fluctuations.
Hence, as an alternative method we also investigated the free expansion of the
condensate after release from the trap. We found that the hole in the annulus
tends to be filled during the expansion, meaning that a 
vortex is clearly
discernible as a hole in the expanded density 
profile. As a consequence
of the repulsive interactions, the density of the expanded 
cloud is relatively uniform near to the center, 
allowing the width of the vortex core to be used as a
measure of the vorticity. 

In contrast, when interactions are switched off during the expansion we find 
that the central density is less uniform, and 
at long times it tends towards the momentum distribution of the condensate. 
Since the momentum distribution can be also probed by Bragg spectroscopy,
we have explored the possibility of using such a measurement to detect vortex 
states. We find
that for thin annuli in the Thomas-Fermi approximation the distribution takes
the form $|\Phi(k)|^2 \propto [J_\nu (kR)]^2$. Fourier transforming the 
stationary solutions of the Gross-Pitaevskii
equation changes the amplitudes, but not the positions, of the peaks,
potentially allowing the minima of an experimentally measured momentum
distribution to be compared to the zeros of the Bessel function in order to
distinguish different $\nu$ states.

Finally, we would like to stress that most of the
results obtained in the thin annulus limit are expected to be
valid for generic ring-shaped condensates, independent of the specific form of
the trapping potential. In particular, near to the minimum a generic
Mexican hat potential can be approximated by a harmonic well in the radial
coordinate, allowing a precise mapping onto the results for our quartic
minus quadratic potential.

\acknowledgments

This work was supported by the Ministero dell'Istruzione, dell'Universit\`a e
della Ricerca.


\begin{thebibliography}{99}
\bibitem{spurious}
 The $j=m=0$ solution is excluded since it does not conserve particle number,
 and is therefore unphysical.
\bibitem{corrections}
 The exact solution of Eq.~(\ref{eq:HD lin no rot}) can be formally written as
 a series expansion in powers of $\eta$, i.e.\ 
 $\omega_{jm}^2=\sum_{k}w_{jmk}\eta^k$, $\delta{S}_{jm}(\zeta)=
 {e}^{im\phi}\sum_{k}p_{jmk}(\zeta)\eta^k$, where the functions $p_{jmk}(\zeta)$
 are polynomials of order $j+k$ and parity $(-1)^{j+k}$. Keeping terms up to
 $\eta^2$, for the most interesting cases $j=0,1$ one has
 $\omega_{0m}^2=m^2\eta^2/6$ and $\omega_{1m}^2=2+(m^2/2-3)\eta^2/5$, with the
 corresponding phase variations
 $\delta{S}_{0m}(\zeta)=[P_0(\zeta)+P_2(\zeta)m^2\eta^2/36]{e}^{im\phi}$ and
 $\delta{S}_{1m}(\zeta)=[P_1(\zeta)-P_2(\zeta)\eta/2+
 (m^2/4+6)P_3(\zeta)\eta^2/25]{e}^{im\phi}$.
\bibitem{perturbation}
 At the leading order in $\eta$ and for small angular velocities, where 
 $\delta{n}_{jm}(\zeta) \sim P_j(\zeta){e}^{im\phi}$, the quadrupole operator
 is $x^2-y^2=[P_0(\zeta)+{\eta}P_1(\zeta)]\cos(2\phi)/(2\lambda)
 =(\delta{n}_{0,+2}+\delta{n}_{0,-2})/(4\lambda)+\mathcal{O}
 (\eta)$, mainly exciting the $(j=0,m=\pm 2)$ modes.
 The ($j=1,m=\pm2$) modes, whose excitation amplitude is proportional
 to $\eta$, are responsible for the small-amplitude 
 high-frequency oscillations visible in
 Fig.~\ref{fig:freq5}(b).
\bibitem{3D expansion}
 Experimentally, one could realize a purely 2D expansion by retaining the 
 confinement in the axial direction. A full 3D expansion, inducing a lowering 
 of the effective 2D coupling constant, would enhance the increase of the 
 healing length, changing the scaling of the vortex core radius. However, the
 qualitative behavior, and in particular the factor of two difference between 
 the vortex radii for $\nu=0$ and $\nu=1$, is expected to be preserved.
\end{thebibliography}
\end{document}